\documentclass[aps,prb,10pt,twocolumn,showpacs,floatfix,superscriptaddress,amsmath,amssymb]{revtex4-1}

\usepackage{graphicx}
\usepackage{dcolumn}
\usepackage{bm}
\usepackage{color}
\usepackage{version}

\newcommand{\ba}{\begin{eqnarray}}
\newcommand{\ea}{\end{eqnarray}}
\def\be{\begin{equation}}
\def\ee{\end{equation}}

\def\bimno{Bi$_3$Mn$_4$O$_{12}$(NO$_3$)}

\excludeversion{details}

\begin{document}

\title{Optimizing configurations for determining the magnetic model based on ab-initio calculations}

\author{J. M.\ Matera}
\email{matera@fisica.unlp.edu.ar}
\affiliation{IFLP - CONICET, Departamento de F\'isica, Universidad Nacional de La Plata,
C.C.\ 67, 1900 La Plata, Argentina.}
\affiliation{Departamento de Ciencias B\'asicas, Facultad de Ingenier\'ia Universidad Nacional de La Plata, La Plata, Argentina}

\author{ C. A.\ Lamas}
\email{lamas@fisica.unlp.edu.ar}
\affiliation{IFLP - CONICET, Departamento de F\'isica, Universidad Nacional de La Plata,
C.C.\ 67, 1900 La Plata, Argentina.}
\affiliation{Departamento de Ciencias B\'asicas, Facultad de Ingenier\'ia Universidad Nacional de La Plata, La Plata, Argentina}

\author{L. A.\ Errico}
\affiliation{IFLP - CONICET, Departamento de F\'isica, Universidad Nacional de La Plata,
C.C.\ 67, 1900 La Plata, Argentina.}

\author{A.\ V.\ Gil Rebaza}
\affiliation{IFLP - CONICET, Departamento de F\'isica, Universidad Nacional de La Plata,
C.C.\ 67, 1900 La Plata, Argentina.}

\author{V. I.\ Fernandez}
\affiliation{IFLP - CONICET, Departamento de F\'isica, Universidad Nacional de La Plata,
C.C.\ 67, 1900 La Plata, Argentina.}




\begin{abstract}
In this paper, it is presented a novel strategy to optimize the determination of magnetic couplings by using ab-initio calculations of the energy. 
This approach allows determining efficiently, in terms of a proposed effective magnetic spin model, an optimal
set of magnetic configurations to be simulated by DFT methods. 
Moreover, a procedure to estimate the values of the coupling constants and their error bounds from the estimated 
energies is proposed. This method, based on Monte Carlo sampling,  takes into account the accuracy of the ab - initio simulations. 
A strategy to refine models reusing previously computed configuration energies is also presented. 
We apply the method to determine a magnetic model for the recently synthesized material \bimno. 
Finally, an open source software that implements and automatizes the whole process is presented.
  
\end{abstract}
\maketitle

\section{Introduction}

The physical description of magnetic degrees of freedom in a broad class of compounds is usually based on the well known Heisenberg-Dirac-van Vleck Hamiltonian
\cite{Auerbach:1994,LM_2011,Balents_Nature,stat-trans,Bishop_2012,lamas-capponi,Fisher_2013}.
In this context, the use of ab-initio techniques of electronic structure to determine the effective values of the exchange couplings ($J_{i}$) is a very useful tool for theoretical physicists. 
This theoretical description, despite its simplicity, allows us to understand the basic ingredients that give rise to a wide variety of magnetic phases
\cite{Cabra_honeycomb_prb,Messio,Honecker2000a,Mosadeq,Mulder,Trumper1,Lamas2015b,Brenig2001a,Chubukov1991,Zhang2014,Oshikawa_2013}.
The successful determination of the magnetic parameters lies in the appropriate balance between the selection of the model and the spin configuration used in the 
ab-initio calculations.
In contrast, the super-exchange pathways via bridging ligands may cause that interactions between sites separated by a long distance to be large, forcing us to take very large unit cells.
The computational cost is greatly increased with the unit cell size and the use of a large number of spin configurations is restrictive.
In order to be able to perform calculations of the Heisenberg exchange coupling constants and determine the minimal model we should be able to detect the set of configurations that gives
us the best determination of our model.
In this paper we present a strategy to determine this set of optimal configurations to be used in the ab-initio estimation of the energy.
In this way we are able to obtain a well conditioned system of linear equations to determine the parameters of the magnetic model.

We apply the algorithms to determine the magnetic couplings corresponding to the compund \bimno\cite{matsuda2010disordered}. In this compound the Mn$^{4+}$ ions form a honeycomb lattice.
Two layers of such honeycomb planar configurations are separated by Bismuth atoms, forming an almost isolated bilayer structure separated by a long distance to the next bilayer structure. 
Furthermore, the magnetic susceptibility data\cite{smirnova2009synthesis} and neutron scattering\cite{matsuda2010disordered} suggests two-dimensional magnetism, so it seems reasonable to model the system with a bilayer structure of  Mn$^{4+}$ ions.
There is some experimental evidence that the interlayer coupling and the first and second neighbors intralayer couplings are the most relevant interactions and could be a strong competition between them\cite{matsuda2010disordered}.

The outline of the paper is the following. In section \ref{sec:method} we state criteria used in determining a suitable strategy to obtain a well conditioned set of equations for the exchange couplings.
This system defines a family of representative models. 
In section \ref{sec:application-bimono} we apply the method to define the family of magnetic models corresponding to the material \bimno . 
In section \ref{sec:conclusions} we present the conclusions and perspectives. Finally in the appendices \ref{sec:svd},\ref{app:appendix-alg} and \ref{app:visualbond} we display
the single value decomposition, the detailed algorithm to optimize the set of magnetic configurations and the visual interfase for the scripts that implement the method.

\section{Coupling constants and  Relevant Configurations}
\label{sec:method}
In this section, we discuss a method based on ab-initio calculation of total energy differences to estimate the coupling constants in an effective magnetic model.

We start by considering a certain atomic lattice involving magnetic atoms. Magnetism in matter is an intrinsically quantum phenomenon, requiring for its description a full quantum
treatment of the electronic degrees of freedom. An exact approach of such problem is computationally unfeasible due to the huge size of its associated Hilbert space. 

The usual strategy to tackle this problem is based on the relative weakness of the magnetic contribution to the energy, compared to those coming from the spatial degrees of freedom.
This allows to approximate the true ground state of the system as a linear combination of those Slater's determinant-like states
$$|\beta\rangle=|\{\varphi_i\},\{s^\beta_i\}\rangle_{SL}=\frac{1}{\sqrt{N}}\left||\varphi_1, s^\beta_1\rangle\ldots |\varphi_N, s^\beta_N \rangle\right|$$
that minimizes $E_{\beta}=\langle \beta |{\bf H}_{\rm full}|\beta\rangle$  for each fixed spin configuration $\{s^\beta_i\}$. By construction, these states define an orthogonal basis of the ground multiplet of the system. The evaluation of these optimizations can be performed in a relatively efficient way by means of Hartree Fock or Density Functional Theory (DFT)-like methods\footnote{In DFT, the test wave functions are not exactly Slater's determinats, but the argument holds: GS associated to different spin configurations are orthogonal among them.}. This approximation is justified if, for each $\{s_i\}$, the corresponding spectrum is gapped. In such a case, the true Ground State (GS) (and its low-lying excited states) can be obtained diagonalizing 
$$
{\bf H}_0= \sum_{\beta\beta'}|\beta\rangle\langle \beta| {\bf H}_{\rm full}|\beta'\rangle  \langle \beta'|
$$ 
DFT and Hartree-Fock formalisms provide a method to evaluate (individual) diagonal elements of ${\bf H}_0$ in an efficient way. Notice that at this point, ${\bf H}_0$ is
still in principle a huge matrix, in a way that even to evaluate every diagonal entry is a non-affordable task. To go further, we will suppose that ${\bf H}_0$ 
can be approximated by a simpler model, depending on a relatively small number of parameters. The family of Heisenberg's models
$${\bf H}_{ eff}[\{J_{\alpha}\}]=\sum_{\alpha=1}^{M} \frac{J_\alpha}{2}\sum_{(i,j)\in B_\alpha} \vec{\bf S}_i\cdot\vec{\bf S}_j+ J_0{\bf 1}$$
supplies a very versatile class of models, with a rich phase space, that do not break $SU(2)$ symmetry. Here, $B_a$ are sets of equivalent pairs of sites in the lattice, $J_\alpha$ the corresponding
coupling constants ( $J_0$ is a global energy offset). The coupling constants $J_{\alpha}$ can then be choosen in a way that ${\bf H}_{ eff}$ has diagonal entries close to the computed diagonal elements on ${\bf H}_0$. Notice that the condition on all the diagonal entries in both Hamiltonians defines an overconditioned set of linear equations for $J_{\alpha}$, in general it will not be possible to find  $J_{\alpha}$ for a perfect match. On the other hand, as energies in DFT can only be estimated up to a finite accuracy $\Delta \varepsilon$, it makes sense to ask whether the diagonal elements in both matrices differ in less than $\Delta \varepsilon$. We define then the set of $\Delta \varepsilon-$\emph{compatible parameters}
\begin{equation}
\label{eq:compatibility}
{\cal C}_{\Delta \varepsilon}:= \{ \{J_\alpha\} / |\langle \beta|{\bf H}_{ eff}[\{J_\alpha\}]|\beta\rangle-E_{\beta}|<\Delta \varepsilon, \;\;\forall  |\{s_i\}\rangle\}\,.
\end{equation}
in a way that any element in ${\cal C}_{\Delta \varepsilon}$ generates a ${\bf H}_{eff}$ with diagonal elements compatible with ${\bf H}_0$ upto the tolerance $\varepsilon$. If ${\cal C}_{\Delta \varepsilon}$ is small enough, we can expect that ${\bf H}_{eff}$ leads  to the same physical predictions for any choice of $\{J_\alpha\}$ in ${\cal C}_{\Delta \varepsilon}$.
Once we have a \emph{representative} choice for $\{J_\alpha\}$, we can deal with ${\bf H}_{eff}$ by means of different analytical and numerical methods\cite{Auerbach:1994}, 
ranging from boson maps\cite{Sachdev1990,Trumper2,Coleman,Oshikawa_2013,Cabra_honeycomb_prb,Trumper1,Mulder,Zhang_PRB_2013,Brenig2016}, path integrals\cite{lamas-capponi,Lamas2015}, exact diagonalization\cite{Honecker2000a}, DMRG\cite{Honecker2000a,Elias2017}, etc.

\begin{figure}
\includegraphics[width=8cm]{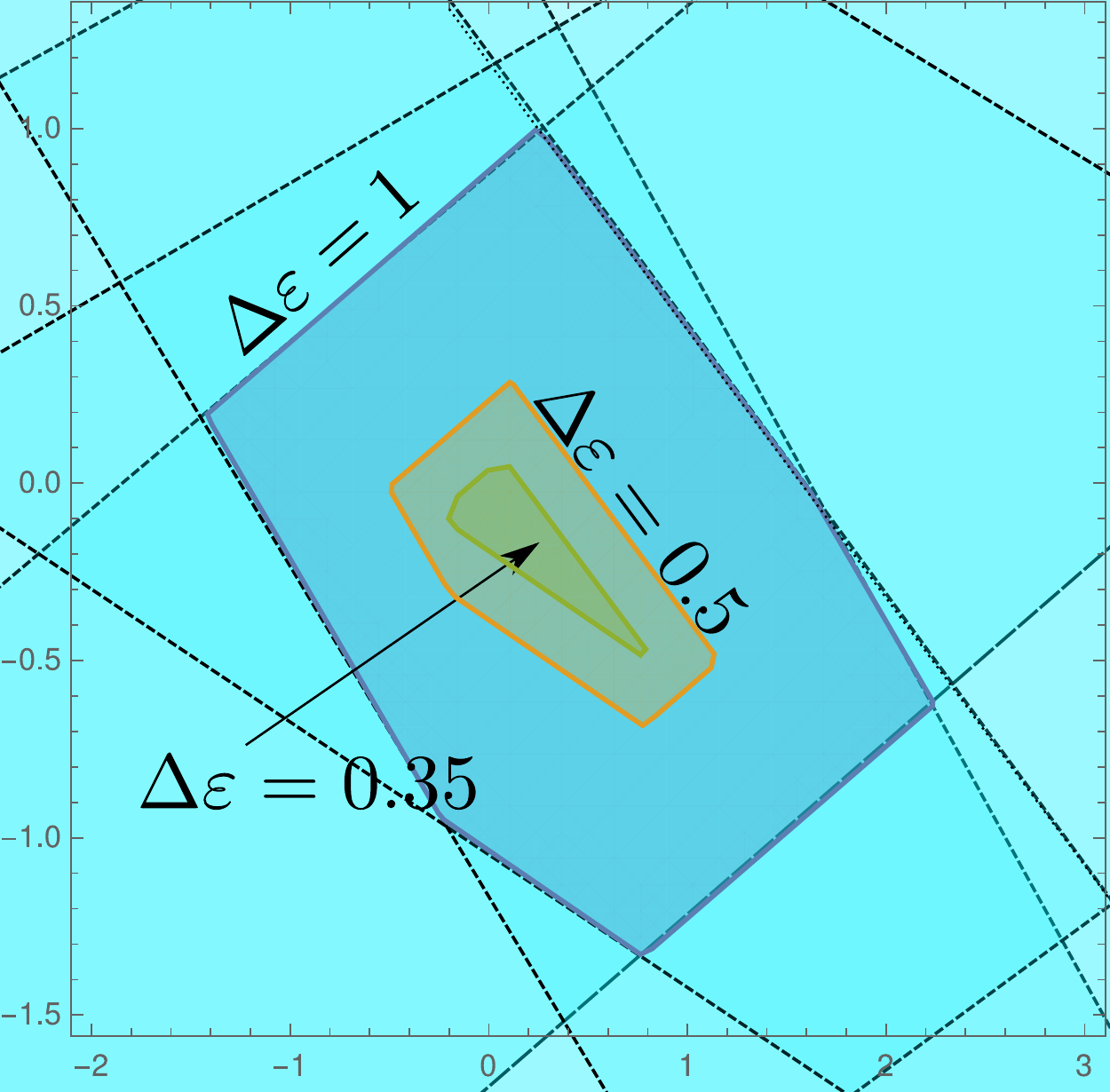}
\caption{Color online. Compabilibity regions for different values of $\Delta \varepsilon$. For some finite value $\Delta \varepsilon_0>0$,
  the compatibility region becomes empty. Dashed lines represent the boundary of the hemi-spaces (semi-planes) defining the largest polytope.}
  \label{fig:compat1}
\end{figure}

If the system involves just a very small number of \emph{magnetic} atoms, the set ${\cal C}_{\Delta \varepsilon}$ can be characterized by the evaluation of the energies of all the possible spin configurations. Since ${\bf H}_{eff}$ is linear in the coupling constants $J_\alpha$, we can rewrite Eq. \ref{eq:compatibility} as
$$
{\cal C}_{\Delta \varepsilon}= \{ \vec{J} / \|A \cdot \vec{J} -\vec{E}_0\|_{\infty}<\Delta \varepsilon \}\,,
$$
where $\vec{J}$ is a vector with components $J_\alpha$, $\vec{E}_0$ is a vector with components $E_{\beta}$, and $A$ is a matrix with coefficients
$$[A]_{\beta \alpha}=\frac{1}{2}\sum_{(i,j)\in B_\alpha}\langle \{s_i\}_\beta|\vec{S}_i\cdot\vec{S}_j |\{s_i\}_\beta\rangle$$ 
and
$$
\|\vec{v}\|_{\infty}=\max_i |\vec{v}_i|
$$
is the \emph{maximum norm} or \emph{infinity norm}\cite{Bhatia97}. As a result, if ${\cal C}_{\Delta \varepsilon}$ is not empty, it is a convex polytope, i.e., a convex set coming from the intersection of many hemi-spaces (see Fig \ref{fig:compat1}).

As the number of magnetic atoms becomes larger (let's say, $>6$), the number of configurations grows exponentially, and therefore the evaluation of all the constraints defining the compatibility polytope ${\cal C}_{\Delta \varepsilon}$ is not feasible anymore. However, we can \emph{bound}
${\cal C}_{\Delta \varepsilon}$ just looking at a smaller set of configurations: defining ${\cal S}_n$ as a subset of $n$ elements from the  set of all possible configurations, we can define
$$
{\cal C}_{\Delta \varepsilon}({\cal S}_n):= \{ \vec{J} / \|A' \cdot \vec{J} -\vec{E}'_0\|_{\infty}<\Delta \varepsilon \} 
$$
where $A'=A'({\cal S}_n)$ and $E_0'=E_0'({\cal S}_n)$ are built in such a way those rows correspoding to the spin configurations in ${\cal S}_n$ are preserved. 

\begin{figure}
\includegraphics[width=8cm]{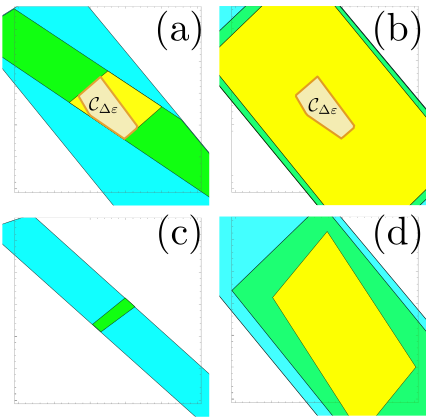}
\caption{Color online. Convergence of the compatibility region with the number of computed configurations. Top: Compatible model. (a) good choice of configurations. (b) a bad one. Bottom: Non compatible model. (c) good choice of configurations. The incompatibility was verified.  (d) bad choice. The incompatibility is not apparent. }
  \label{fig:convergence}
\end{figure}

These sets satisfies ${\cal C}_{\Delta \varepsilon}({\cal S}_n)\subset {\cal C}_{\Delta \varepsilon}({\cal S}_{n'})$ if ${\cal S}_{n'}\subset {\cal S}_{n}$
and ${\cal C}_{\Delta \varepsilon}\subset {\cal C}_{\Delta \varepsilon}({\cal S}_n)$, so that as we increase the number of evaluated configurations, the compatibility set does not increase its size.
Typically, assuming that the spectrum of ${\bf H}_{eff}$ can be accurately represented by a spin Hamiltonian with a moderately small number of couplings, most of the configurations provide no information or just redundant information. In this way, a very tight bound for  ${\cal C}_{\Delta \varepsilon}$ can be obtained by just considering a small set of configurations, with a size of the order of the number of free parameters in the model.
For this, it is crucial to pick the set of configurations in an optimal way: otherwise, ${\cal C}_{\Delta \varepsilon}({\cal S}_n)$ can stay covering a much larger region than ${\cal C}_{\Delta \varepsilon}$, even for quite large $n$ (see Fig \ref{fig:convergence}).

\subsection{Choosing an optimal set of configurations}

One problem about optimizing the size of ${\cal C}_{\Delta \varepsilon}({\cal S}_n)$ is that, in order to be evaluated, it is necesary to know the energy of each configuration, which in general requires a lot of computing time. One strategy to avoid this issue is to make use of the inequality
\begin{equation}
\frac{\|\vec{x}\|_\infty}{\sqrt{n}} \leq  \frac{\|\vec{x}\|_2}{\sqrt{n}} \leq \|\vec{x}\|_\infty \leq \|\vec{x}\|_2
  \label{eq:metricbound}
\end{equation}
being $\|\vec{x}\|_2=\sqrt{\sum_{i=1}^n \vec{x}_i^2}$ the \emph{euclidean norm} and $n$ the number of components of the vector\cite{Bhatia97}. With this in mind, we define

\begin{equation}
\label{eq:compatibility2}
{\cal C}^{(2)}_{\Delta \varepsilon}({\cal S}_n):= \{ \vec{J} / \|A' \cdot \vec{J} -\vec{E}'_0\|_{2}<\Delta \varepsilon \}\,,
\end{equation}
where again, $A'=A'({\cal S}_n)$ and $E_0'=E_0'({\cal S}_n)$ are the restrictions of $A$ and $\vec{E}_0$ to the rows associated to ${\cal S}_n$. These sets define ellipsoids centered at the minimum of the quadratic form
\begin{equation}
  \chi^2(\vec{J})= \|A'.\vec{J}-\vec{E}_0'\|^2
   \label{eq:chi}
\end{equation}
and with main axes defined by the Singular Value Decomposition (SVD) of $A'$:
$$
A'=U \Sigma V^t
$$
where $U^tU=V^tV ={\bf 1}_{M+1}$ and $\Sigma={\rm diag}(\sigma(A')) \in \mathbb{R}^{(M+1) \times (M+1)}$ a diagonal rectangular matrix with the singular values of the matrix $A'$.  The size of ${\cal C}^{(2)}_{\Delta \varepsilon}({\cal S}_n)$ is then bound by

$$
\|{\cal C}_{\Delta \varepsilon}^{(2)}({\cal S}_n)\| < \Delta \varepsilon \;cn({\cal S}_n)
$$

where $cn$ is the \emph{condition number} of $A'$, i.e,
$$
cn(A')=\max_{s \in \sigma(A')}\frac{1}{s} 
$$
in a way that the size of the set depends not on $\vec{E}_0'$ but just on $\vec{A}'$. This allows us to evaluate it \emph{before} any DFT/ab-initio expensive simulation. 

\begin{figure}
\includegraphics[width=8cm]{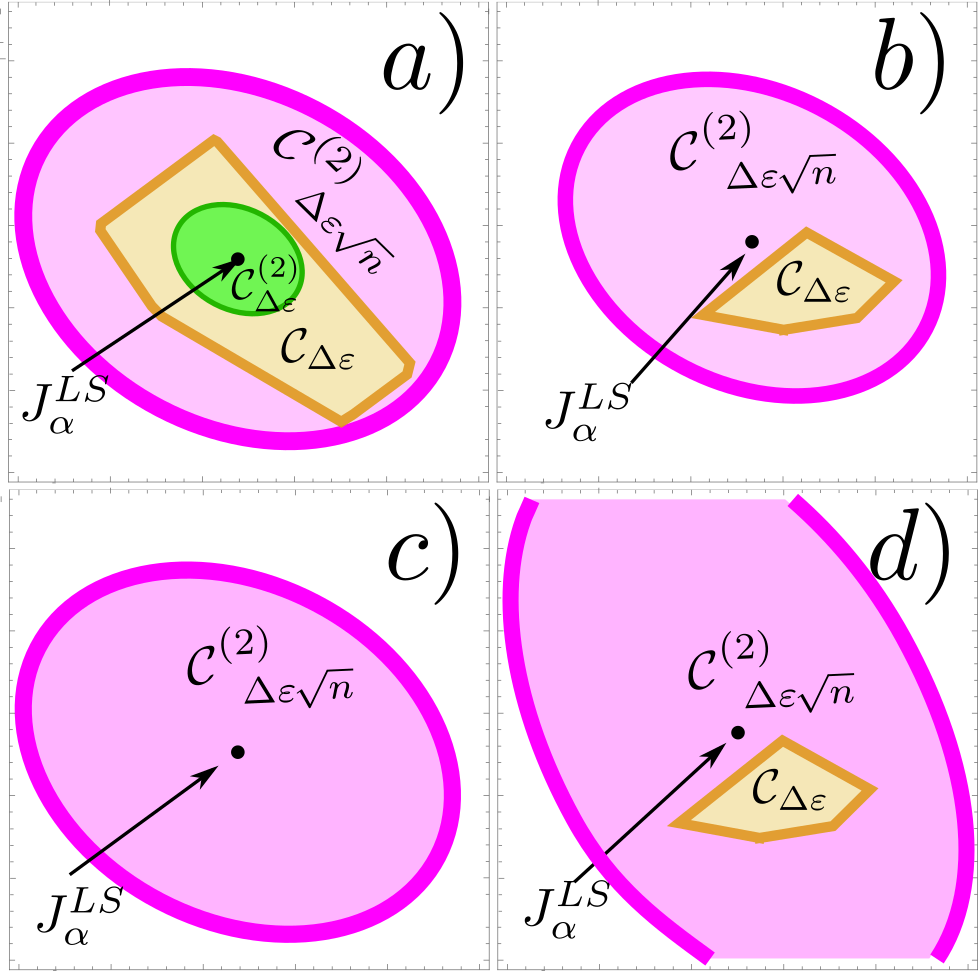}
\caption{Color online. Ellipsoid bound for ${\cal C}_{\Delta \varepsilon}({\cal S}_n)$. a) The inclusion relation among the three sets. b) If the tolerance is reduced, ${\cal C}^{(2)}_{\Delta \varepsilon}({\cal S}_n)$ becomes empty, but the system is still compatible. In this case, $J_\alpha^{LS}$ does not belong to the compatibility zone. c) If the tolerance is reduced further, the model becomes incompatible, but ${\cal C}_{\Delta \varepsilon \sqrt{n}}({\cal S}_n)$ still is non empty. d) For fix tolerance, as the number of configuration grows, ${\cal C}^{(2)}_{\Delta \varepsilon}({\cal S}_n)$ becomes empty,
while ${\cal C}^{(2)}_{\Delta \varepsilon \sqrt{n}}({\cal S}_n)$ grows. }
  \label{fig:convergence2}
\end{figure}

The usefulness of Def. (\ref{eq:compatibility2}) to bound the size of (\ref{eq:compatibility}) comes from Eq. \ref{eq:metricbound}, that leads up to
$$
{\cal C}^{(2)}_{\Delta \varepsilon}({\cal S}_n) \subset {\cal C}_{\Delta \varepsilon}({\cal S}_n) \subset {\cal C}^{(2)}_{\Delta \varepsilon \sqrt{n}}({\cal S}_n)
$$
in a way that ${\cal C}_{\Delta \varepsilon}\neq \O$  if ${\cal C}^{(2)}_{\Delta \varepsilon}({\cal S}_n)\neq \O$, and
${\cal C}^{(2)}_{\Delta \varepsilon\sqrt{n}}({\cal S}_n)$ bounds ${\cal C}_{\Delta \varepsilon\sqrt{n}}({\cal S}_n)$ (see Fig \ref{fig:convergence2}).

From the previous discussion, the strategy to obtain a good set of configurations to bound the compatibility zone is to look for ${\cal S}_n$ 
that minimizes the cost function
\begin{equation}
C({\cal S}_n):= \sqrt{n}\,cn(A'({\cal S}_n))
  \label{eq:costfunction}
\end{equation}

Formally, the problem of finding the absolute minimum of $C({\cal S}_n)$ is hard, since it typically presents many relative minima with
approximately the same cost. However, what we actually need is just one these relative minima, which can be efficiently achieved by the algorithm
presented in the appendix \ref{app:appendix-alg}.

\subsection{Estimation of  $J_{\alpha}$ and its uncertancies}

Once an optimal set of configurations is determined, the corresponding magnetic energies can be estimated by means of DFT simulations. The next step is then to find the representative value $J^{(0)}_{\alpha}$ according to them. In the standard approach, $J^{(0)}_{\alpha}$ is estimated by
$$J_\alpha^{LS}={\rm argmin}_{J_\alpha} \chi^2(J_{\alpha})$$
the least square condition. This approach is valid if ${\cal C}_{\Delta \varepsilon}^2({\cal S}_n)\neq \O$ for the considered tolerance, since in that case $J^0_{\alpha}$ belongs to ${\cal C}_{\Delta \varepsilon}({\cal S}_n)$. This can always be achieved by choosing a large enough value for tolerance. However, in that case, the uncertancy in the estimated coupling constants could result overestimated regarding the true accuracy of the ab-initio simulation. This is a problem because the accuracy of the simulations are usually close to the scale of energy of the coupling constants which are being calculated. As a result, the estimated values for $J_{\alpha}$ are smaller in magnitud than the uncertancies, in a way that at the end of the day we are not able to state even the sign of the couplings.

\begin{figure}
\includegraphics[width=8cm]{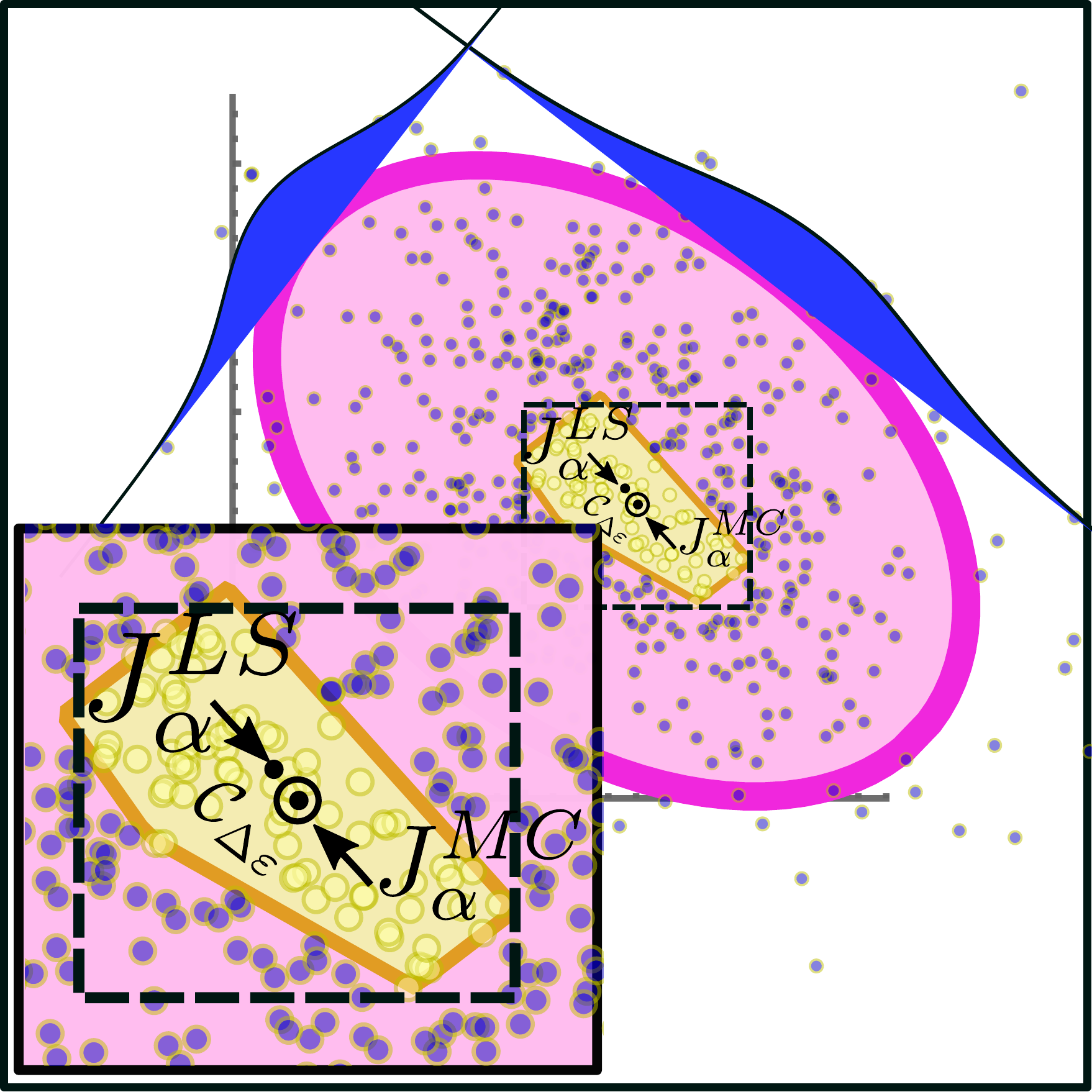}
\caption{Color online. Monte Carlo estimation of the bounds for $J_{\alpha}$. Points are spread around the center of the ellipsoid with a gaussian distribution having a correlation matrix  proportional to $A'^t A'$. The compatibility region is sampled by keeping those points that belongs to it.
  The dashed box approximately bounds the region of the compatibility region.}
  \label{fig:mcdeterm}
\end{figure}

To get a more realistic estimation, a Monte Carlo sampling of the region ${\cal C}_{\Delta \varepsilon}({\cal S}_n)$ can be performed in order to get the limit values of the compatible $J_{\alpha}$. An efficient stragegy consists on explore a set of random points with a gaussian distribution around $J^{LS}_{\alpha}$, with a correlation matrix $\lambda (A^t A)^{-1}$ (see Fig. \ref{fig:mcdeterm}).

\subsection{Improving a model}

If the accuracy in the DFT energy estimation is high enough, it could happen that the proposed model becomes incompatible. In that case, a more sofisticate model can be required, for instance, by considering different two sets of coupligs that initially were considered with the same value, or by adding interactions with more distant neighbors. For this new model, the optimal set of configurations to determine the new coupling constants can be different. Although, it can be expected that there is a set of configurations, including those in the optimal set for the simpler model, that is also a relative minimum. This allows us to reuse the energies obtained in the simpler model, reducing the computational cost of estimating couplings in the new model. A way to achieve this consists on, first, optimize $C({\cal S}_{n}\cup {\cal S}'_{n'} )$ regarding an ${\cal S}_{n'}$ with a fixed $n'$, being ${\cal S}_n$ the optimal set for the simpler model.
Due to the factor $\sqrt{n}$ in $C({\cal S}_n)$, this result can be improved by reducing the size of ${\cal S}_n\cup {\cal S}'_{n'}$, which can be accomplish by dropping \emph{one by one} elements from the set.

\subsection{Visualbond Spectrojotometer}

The method proposed here is suitable for being automatized. In this aim, an open source project has been developed\cite{visualbond} to provide to the community a software tool that performs each step in the analysis. The tool consists on a set of python libraries that helps to build magnetic models from structural data of the target magnetic compound (in cif format), and once the model is defined, to determine the optimal magnetic configurations to be evaluated, and finally, when the ab-initio simulations are performed, to estimate the corresponding coupling constants including their error bounds.

\section{Aplication: Magnetic model for \bimno}
\label{sec:application-bimono}

The synthesis of the material bismuth oxynitrate, {\bimno}, obtained by Smirnova
\textit{et al.}\cite{smirnova2009synthesis}, has given a great impulse to the study of two dimensional effective antiferromagnetic models on the bilayer honeycomb lattice.
Here, the Mn$^{4+}$ ions form a honeycomb lattice, grouped in pairs and  separated by  a large distance by bismuth atoms.
For this material the bilayer honeycomb lattice brings an appropriate geometry to build an effective Hamiltonian capable to describe its magnetic properties.
The Heisenberg model on the bilayer Honeycomb lattice has been attracted a plenty of theoretical studies in the last years\cite{Fisher_2013,Arlego201415,Cabra_honeycomb_prb,Cabra_honeycomb_2,fouet1975investigation,kandpal2011calculation,Richter2017,Mosadeq,Oitmaa_2012}.
The richness of this model makes it very interesting from the theoretical point of view.   

This two dimensional nature of the effective model is reinforced by magnetic susceptibility data.
Also, no evidence of long-range ordering has been observed down to 0.4 K\cite{smirnova2009synthesis,matsuda2010disordered,okubo2010high,azuma2011frustrated}.
On the other hand, recent experimental studies\cite{matsuda2010disordered} have suggested  that the inter-layer coupling,
as well as on-layer nearest and, to a lesser extent, next-nearest couplings, are dominant.
This last characteristic of the model makes the ab-initio calculation a good tool to understand the mechanism involved by determining an appropriated
effective model.

Recently\cite{Alaei.2017} the magnetic model of \bimno\ has been estimated by using fifty-four different spin configurations and determined the energy by ab-initio calculations with an error $\Delta E= 0.5$ meV.
As the DMFT calculation of the energy is hard, it is convenient to reduce the number of spin configurations needed to obtain the magnetic couplings. Using the strategy presented in the previous sections we determine a set of  
optimal configurations and calculate the magnetic couplings.
In table \ref{tab:jotas} we present the couplings obtained in\cite{Alaei.2017} using fifty-four calculations of the energy and the result using our optimal eleven configurations. All the couplings agree up to the error. 
Using our strategy to select the optimal configurations before the ab-initio determination of energy can save many hours of machine work and allows us to work with larger unit cells.

\begin{table}
\begin{center}
\begin{tabular}{|c|c|c|}
\hline
  $J_{i}/|J_{1}|$ & $\mbox{11 configurations}\atop{(|J_1|=0.346{\rm meV})}$ & $\mbox{54 configurations}\atop{(|J_1|=0.349{\rm meV})}$ \\ \hline
$J_{1}/|J_{1}|$ & $-1.0 \pm 0.1$& $-1.0 \pm 0.06$\\ \hline
$J_{2}/|J_{1}|$ & $-0.12 \pm 0.06$& $-0.11 \pm 0.04$\\ \hline
$J_{3}/|J_{1}|$ & $-0.09 \pm 0.07$& $-0.09 \pm 0.05$\\ \hline
$J_{0}/|J_{1}|$ & $-0.3 \pm 0.21$& $-0.3 \pm 0.12$\\ \hline
$J_{1c}/|J_{1}|$ & $-0.1 \pm 0.09$& $-0.11 \pm 0.06$\\ \hline
$J_{2c}/|J_{1}|$ & $-0.05  \pm 0.07$& $-0.03 \pm 0.04$\\ \hline
$J_{3c}/|J_{1}|$ & $-0.07 \pm 0.08$& $-0.06 \pm 0.05$\\ \hline
\end{tabular}
\caption{\label{tab:jotas}
Coupling constants obtained by ab-initio calculations with fifty-four configurations in ref\cite{Alaei.2017}
}
\end{center}
\end{table}

\begin{figure}[t!]
\begin{center}
\includegraphics[width=0.9\columnwidth]{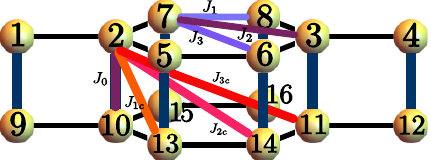}
\caption{Labeling for the Mn atoms in the unit cell and Coupling constant considered in the model.}
\label{fig:numeracion}
\end{center}
\end{figure}

The optimal subset of eleven configurations that we have found with the method developed in Sec \ref{sec:method} is
\ba
|1\rangle &=& | \uparrow\; \uparrow\; \uparrow\; \uparrow\; \uparrow\; \uparrow\; \uparrow\; \uparrow\; \uparrow\; \uparrow\; \uparrow\; \uparrow\; \uparrow\; \uparrow\; \uparrow\; \uparrow\;\rangle \\ \nonumber
|2\rangle &=& | \uparrow\; \uparrow\; \downarrow\; \downarrow\; \uparrow\; \downarrow\; \uparrow\; \downarrow\; \uparrow\; \uparrow\; \downarrow\; \downarrow\; \uparrow\; \downarrow\; \uparrow\; \downarrow\;\rangle \\ \nonumber
|10\rangle &=& | \uparrow\; \downarrow\; \uparrow\; \downarrow\; \uparrow\; \downarrow\; \uparrow\; \downarrow\; \downarrow\; \uparrow\; \downarrow\; \uparrow\; \downarrow\; \uparrow\; \downarrow\; \uparrow\;\rangle \\ \nonumber
|17\rangle &=& | \downarrow\; \downarrow\; \downarrow\; \downarrow\; \uparrow\; \uparrow\; \uparrow\; \uparrow\; \downarrow\; \downarrow\; \uparrow\; \uparrow\; \uparrow\; \uparrow\; \uparrow\; \uparrow\;\rangle \\ \nonumber
|18\rangle &=& | \downarrow\; \downarrow\; \downarrow\; \downarrow\; \uparrow\; \uparrow\; \uparrow\; \uparrow\; \downarrow\; \downarrow\; \downarrow\; \downarrow\; \uparrow\; \uparrow\; \uparrow\; \uparrow\;\rangle \\ \nonumber
|24\rangle &=& | \uparrow\; \uparrow\; \downarrow\; \uparrow\; \uparrow\; \uparrow\; \downarrow\; \uparrow\; \downarrow\; \uparrow\; \uparrow\; \downarrow\; \uparrow\; \uparrow\; \uparrow\; \downarrow\;\rangle \\ \nonumber
|28\rangle &=& | \uparrow\; \downarrow\; \uparrow\; \downarrow\; \uparrow\; \downarrow\; \uparrow\; \downarrow\; \uparrow\; \downarrow\; \uparrow\; \downarrow\; \uparrow\; \downarrow\; \uparrow\; \downarrow\;\rangle \\ \nonumber
|34\rangle &=& | \downarrow\; \downarrow\; \uparrow\; \downarrow\; \uparrow\; \uparrow\; \downarrow\; \downarrow\; \downarrow\; \downarrow\; \downarrow\; \uparrow\; \uparrow\; \downarrow\; \uparrow\; \uparrow\;\rangle \\ \nonumber
|41\rangle &=& | \downarrow\; \uparrow\; \uparrow\; \uparrow\; \uparrow\; \downarrow\; \downarrow\; \downarrow\; \uparrow\; \downarrow\; \uparrow\; \downarrow\; \downarrow\; \downarrow\; \uparrow\; \uparrow\;\rangle \\ \nonumber
|45\rangle &=& | \downarrow\; \uparrow\; \uparrow\; \downarrow\; \downarrow\; \uparrow\; \downarrow\; \downarrow\; \downarrow\; \downarrow\; \downarrow\; \downarrow\; \uparrow\; \uparrow\; \uparrow\; \uparrow\;\rangle \\ \nonumber
|47\rangle &=& | \uparrow\; \uparrow\; \uparrow\; \uparrow\; \uparrow\; \uparrow\; \uparrow\; \uparrow\; \downarrow\; \downarrow\; \downarrow\; \downarrow\; \downarrow\; \downarrow\; \downarrow\; \downarrow\;\rangle
\ea
where the configurations are labeled according to those presented in ref. \onlinecite{Alaei.2017} and sites are labeled as in Fig. \ref{fig:numeracion}.

\section{Conclusions}
\label{sec:conclusions}

In this paper we have presented a method to find an optimal set of configurations in order to determine the couplings in a magnetic model by means of
ab-initio calculations. This strategy allows to enhance the use of ab-initio calculations to establish the parameters of an effective magnetic model to be consistent with
the calculation of the energies. We apply the method to calculate the family of coupling constants consistent with the ab-initio energies. We show that, while taking an optimal set of magnetic configurations, it is possible to reduce 
the number of ab-initio calculations to determine the model with a given error and we obtain results for the coupling constants in agreement to previous calculations. 
Finally we make available a free software to implement the algorithms described in the present work. 

\section*{Acknowledgments}
This research was partially supported by CONICET (Grant No. PIP 0691, PIP0747, PIP0720 and P-UE 22920170100066CO), UNLP (Grant No. 11/X678, 11/X680, 11/X708, 11/X788, 11/X792, 11/X806), ANPCyT (Grant No. PICT 2012-1724, 2013-2616, 2016-4083), UNNOBA (Grant No. SIB0176/2017), and Proyecto Acelerado de C\'alculo 2017 de la Red Nacional de Computaci\'on de Alto Desempe\~no (SNCAD-MINCyT) - HPC Cluster, Rosario - Argentina.

\appendix

\section{Singular Value Decomposition (SVD)}
\label{sec:svd}

Let $A\in {\mathbb C}^{n\times m}$ be a rectangular matrix  with complex entries. Then, the theorem of the Singular Value Decomposition states that there exist two rectangular matrices $U \in {\mathbb C}^{n \times k}$, $V \in {\mathbb C}^{m \times k}$ with $k\leq m,n$, and a diagonal positive matrix
${\Sigma}\in {\mathbb C}^{k\times k}$ such that
\begin{eqnarray}
  A&=&U \Sigma V^\dagger\\
  U^\dagger U&=&V^\dagger V={\bf 1}_{k,k}
\end{eqnarray}
in a way that the columns in $U$ ($V$) are \emph{orthonormals}  (see, for instance,\cite{Bhatia97}). 
The diagonal elements of $\Sigma$ are called the singular values of $A$, denoted by $\sigma(A)={\rm diag}(\Sigma)={\sigma_1,\ldots \sigma_k}$, and the corresponding columns in $U$ ($V$) the left (right) singular vectors.  

Since $A A^\dagger=U \Sigma^2 U^\dagger$ and $A^\dagger A=V \Sigma^2 V^\dagger$  are two  semi-definite positive matrices, the left (right) singular vectors of a matrix correspond to the eigenvectors of  $A A^\dagger$ ($A^\dagger A$) with non vanishing eigenvalues. If $A\in \mathbb{R}^{n\times m}$,
$U$ and $V$ can be choosen as real matrices.

Among many other applications, the singular value decomposition allows to solve the linear least square problem
$$
\vec{x}_0={\rm arg min}_{\vec{x}\in {\mathbb R}^n}\|A\cdot \vec{x} - \vec{b}\|^2
$$
given $A\in \mathbb{C}^{m\times n}$ ($A\in \mathbb{R}^{m\times n}$) and $\vec{b}\in \mathbb{C}^m$ ($\vec{b}\in \mathbb{R}^m$ ), and where $\| \ldots \|\equiv \| \ldots \|_2$ stands here for the Euclidean norm. To solve this, the \emph{Moore-Penrose} pseudo inverse is introduced 
$$
A^+=(A^tA)^{-1}A^{t}=A^{t}(A A^t)^{-1}  =V \Sigma^{-1}  U^t
$$
satisfying
\begin{eqnarray}
  A A^+&=&U \Sigma V^t V \Sigma^{-1}U^{t}=U U^{t}=\Pi_L\\
  A^+ A&=&V \Sigma^{-1} U^t U \Sigma V^{t}=V  V^{t}=\Pi_R
\end{eqnarray}
being $\Pi_L$ ($\Pi_R$) a \emph{projector} over the subspace generated by the left (right) singular vectors of $A$. Now, choosing 
$\vec{x}_0= A^+ \vec{b}$ we obtain
\begin{eqnarray*}
  \|A(\delta\vec{x}+\vec{x}_0)-\vec{b}\|^2&=& \|A\delta\vec{x}+ A A^+\vec{b}-\vec{b}\|^2\\
                                        &=&\|\Pi_L A\delta\vec{x}\|^2+ \|({\bf 1}_{m\times m}-\Pi_L)\vec{b}\|^2\\
  &\geq& \|({\bf 1}_{m\times m}-\Pi_L)\vec{b}\|^2
\end{eqnarray*}

In the same way, it follows that the region
$$
\|A\cdot \vec{x} - \vec{b}\| < \varepsilon
$$
(if it is not empty) corresponds to an ellipsoid center at $\vec{x}_0=A^+ \vec{b}$ with main $i$-esim axis parallel to the
directions of the $i$-esim left singular vectors, with a size $d_i=\frac{\sqrt{\varepsilon^2-|{A}\vec{x}_0-\vec{b}|^2}}{\sigma_i}$.

\section{The algorithm}
\label{app:appendix-alg}

A first problem to find an optimal set of equations is related to how to produce a set of inequivalent spin configurations. This is important since for a system large enough, symmetries present on it imply that two different spin configurations result in the same condition over the couplings. To check the equivalence under symmetries is non trivial for general lattices. A more direct way to reject equivalent configurations is just by comparing the coefficients of the linear equations associated to them. The algorithm is then

\begin{enumerate}
\item   Pick a random integer number between $0$ and $2^{N-1}$, and assign a spin configuration according to its binary digits.
\item   Build the coefficients $A_{ik}$ from this configuration.
 \item  If the size of the set of configurations is of the right size, return the set.
  \item Pick a new random configuration in the same way than 1, and add it to the set of configurations.
 \item Build $A_{ik}$. If the last row has the same coefficients that another row, remove the added configuration from the set
   and go to 4. Otherwise, go to 3.
\end{enumerate}

The same routine can be used if we want to enlarge a previous set of independent configurations. 

The second subrutine that we need for the optimization consists on a method for given a set of configurations, look for a smaller set, with a fixed size  or not, in a way that the cost function is maximized. To perform this task, we implement the following algorithm:

\begin{enumerate}
\item  Build the matrix $A_{ik}$.
\item  Compute its SVD $A_{ik}=U \Sigma V^t$.
\item  Assign weights to the configurations according to the (Euclidean) norm of the corresponding row in $U$.
\item  Take the first $L\geq M$ configurations with larger weight.
\item  If the set leads to a singular set of equations, new configurations are added one by one until the full set is saturated. 
\end{enumerate}

If we want to obtain a set of optimal size, we start with this method picking $L=M$ configurations in the step 4. Then, in the step 5, we continue adding configurations until the cost function starts to grow. In both cases, it is convenient to decouple the determination of the coupling constants from the determination of the non magnetic energy. This can be achieved by adding a row to ${A}_{ik}$ built by averaging the rows associated to each configuration, and afterwards, substracting it to the other rows.

To find out the set of $N>M$ optimal magnetic configurations for a given model, the algorithm starts from a random set of configuration, of size larger than the number $N'>N$ of configurations to be determined. To generate it, we pick integer random numbers between 0 and $2^{l-1}$, and assign
the spin orientations according to its binary digits. From this set, the matrix $A[{\cal S}]$ is built. If two configurations $i$ and $j$ are equivalent, in a way that $A[{\cal S}]_{ik}=A[{\cal S}]_{jk} \forall_k$, the configuration $j$ is dropped and a new configuration is picked. 
Since the last column in $A[{\cal S}]$, corresponding to the coefficient of the energy offset has all its elements equal to 1, the matrix is \emph{normalized} by averaging all the rows, substracting it to each row, and adding the average row at the end. This procedure leads to an equivalent set of equations, but where the equation for the coupling constants are \emph{decoupled} from the equation for the non-magnetic contribution $E_0$. From this initial set, the corresponding SVD 
$$
A[{\cal S}]= U \Sigma V^t
$$
and the cost function \ref{eq:costfunction} are evaluated. If the cost function is $\infty$, then a new set of random inequivalent configurations is added to the set, and the procedure is repeated until a finite cost function is achieved, or a maximal size is attained. In the last case, the algorithm fails.
Once an initial set of configurations with finite cost function is found, a weight to each configuration is assigned according the (euclidean) norm of the i-esim column of $U$. Using this weights, the subset of $N$ configurations of larger weights is kept. From these configurations, the normalized matrix $A[{\cal S}]$ is evaluated, and from it, the new cost function.

\section{Visualbond: graphical environment}

\label{app:visualbond}

In this section we are going to discuss the basics for the use of VisualBond, the graphic interface for the spectrojotometer
library\cite{visualbond}. A more detailed description of functionalities and a set of examples can be found in the application documentation.

The graphic interface is composed by three tabs, corresponding to each step of the process, a menu and a status region.

The process starts by providing structural data for the compound to be simulated to the application. The current version accepts CIF files for this input, as well as
 partial support for struct files (Wien2k). Once the model is loaded, a reduced .cif file is shown in the edition box of the first tab. 
The reduced CIF includes just the lattice parameters, the set of magnetic atoms with their positions, and the set of magnetic bonds.
In the first tab, at the left of the edition box, there is  a dialog for automatically add new bonds in terms of the interatomic distances. 
By modifying the CIF file we can define different bonds among magnetic atoms.

Once the model is loaded, the other two tabs become enabled. We can then use the second tab for looking for a set of optimal configurations,
loading a previously found set, or adding by hand a particular set.
Once the set of configurations is loaded, the corresponding set of equations can be calculated and shown in different formats. 
Notice that by default, the energy associated to each configuration is established to ``nan''.

Once the simulations have been performed, these ``nan'' values must be replaced by the energies obtained from the simulations.
After doing that, in the third tab, the configurations edit box is found again. 
At its left, we can set  the parameters for the fitting - the tolerance in the computed energies, the method for bounding the compatibility zone and its parameters - 
and a button to build the estimation of the coupling constants.
At the right side, the corresponding equations and the estimation of the parameters, and the estimation of the relative discrepances between the values of the
energy from the simulation and from the magnetic model, can be set. This tolerances can be also presented as a plot, helping to identify which configurations are best
represented by the model.

\bibliographystyle{aipnum4-1}
\bibliography{refs-J1-J2-Jp}

\begin{thebibliography}{39}%
\makeatletter
\providecommand \@ifxundefined [1]{%
 \@ifx{#1\undefined}
}%
\providecommand \@ifnum [1]{%
 \ifnum #1\expandafter \@firstoftwo
 \else \expandafter \@secondoftwo
 \fi
}%
\providecommand \@ifx [1]{%
 \ifx #1\expandafter \@firstoftwo
 \else \expandafter \@secondoftwo
 \fi
}%
\providecommand \natexlab [1]{#1}%
\providecommand \enquote  [1]{``#1''}%
\providecommand \bibnamefont  [1]{#1}%
\providecommand \bibfnamefont [1]{#1}%
\providecommand \citenamefont [1]{#1}%
\providecommand \href@noop [0]{\@secondoftwo}%
\providecommand \href [0]{\begingroup \@sanitize@url \@href}%
\providecommand \@href[1]{\@@startlink{#1}\@@href}%
\providecommand \@@href[1]{\endgroup#1\@@endlink}%
\providecommand \@sanitize@url [0]{\catcode `\\12\catcode `\$12\catcode
  `\&12\catcode `\#12\catcode `\^12\catcode `\_12\catcode `\%12\relax}%
\providecommand \@@startlink[1]{}%
\providecommand \@@endlink[0]{}%
\providecommand \url  [0]{\begingroup\@sanitize@url \@url }%
\providecommand \@url [1]{\endgroup\@href {#1}{\urlprefix }}%
\providecommand \urlprefix  [0]{URL }%
\providecommand \Eprint [0]{\href }%
\providecommand \doibase [0]{http://dx.doi.org/}%
\providecommand \selectlanguage [0]{\@gobble}%
\providecommand \bibinfo  [0]{\@secondoftwo}%
\providecommand \bibfield  [0]{\@secondoftwo}%
\providecommand \translation [1]{[#1]}%
\providecommand \BibitemOpen [0]{}%
\providecommand \bibitemStop [0]{}%
\providecommand \bibitemNoStop [0]{.\EOS\space}%
\providecommand \EOS [0]{\spacefactor3000\relax}%
\providecommand \BibitemShut  [1]{\csname bibitem#1\endcsname}%
\let\auto@bib@innerbib\@empty
\bibitem [{\citenamefont {Auerbach}(1994)}]{Auerbach:1994}%
  \BibitemOpen
  \bibfield  {author} {\bibinfo {author} {\bibfnamefont {A.}~\bibnamefont
  {Auerbach}},\ }\href@noop {} {\emph {\bibinfo {title} {Interacting Electrons
  and Quantum Magnetism}}}\ (\bibinfo  {publisher} {Springer-Verlag},\ \bibinfo
  {address} {New York},\ \bibinfo {year} {1994})\BibitemShut {NoStop}%
\bibitem [{\citenamefont {Lhuillier}\ and\ \citenamefont
  {Misguich}(2011)}]{LM_2011}%
  \BibitemOpen
  \bibfield  {author} {\bibinfo {author} {\bibfnamefont {C.}~\bibnamefont
  {Lhuillier}}\ and\ \bibinfo {author} {\bibfnamefont {G.}~\bibnamefont
  {Misguich}},\ }in\ \href@noop {} {\emph {\bibinfo {booktitle} {Introduction
  to Frustrated Magnetism}}}\ (\bibinfo  {publisher} {Springer-Verlag},\
  \bibinfo {year} {2011})\BibitemShut {NoStop}%
\bibitem [{\citenamefont {Balents}(2010)}]{Balents_Nature}%
  \BibitemOpen
  \bibfield  {author} {\bibinfo {author} {\bibfnamefont {L.}~\bibnamefont
  {Balents}},\ }\href@noop {} {\bibfield  {journal} {\bibinfo  {journal}
  {Nature}\ }\textbf {\bibinfo {volume} {464}},\ \bibinfo {pages} {199}
  (\bibinfo {year} {2010})}\BibitemShut {NoStop}%
\bibitem [{\citenamefont {Lamas}\ \emph {et~al.}(2012)\citenamefont {Lamas},
  \citenamefont {Ralko}, \citenamefont {Cabra}, \citenamefont {Poilblanc},\
  and\ \citenamefont {Pujol}}]{stat-trans}%
  \BibitemOpen
  \bibfield  {author} {\bibinfo {author} {\bibfnamefont {C.~A.}\ \bibnamefont
  {Lamas}}, \bibinfo {author} {\bibfnamefont {A.}~\bibnamefont {Ralko}},
  \bibinfo {author} {\bibfnamefont {D.~C.}\ \bibnamefont {Cabra}}, \bibinfo
  {author} {\bibfnamefont {D.}~\bibnamefont {Poilblanc}}, \ and\ \bibinfo
  {author} {\bibfnamefont {P.}~\bibnamefont {Pujol}},\ }\href {\doibase
  10.1103/PhysRevLett.109.016403} {\bibfield  {journal} {\bibinfo  {journal}
  {Phys. Rev. Lett.}\ }\textbf {\bibinfo {volume} {109}},\ \bibinfo {pages}
  {016403} (\bibinfo {year} {2012})}\BibitemShut {NoStop}%
\bibitem [{\citenamefont {Bishop}\ \emph {et~al.}(2012)\citenamefont {Bishop},
  \citenamefont {Li.}, \citenamefont {Farnell},\ and\ \citenamefont
  {Campbell}}]{Bishop_2012}%
  \BibitemOpen
  \bibfield  {author} {\bibinfo {author} {\bibfnamefont {R.~F.}\ \bibnamefont
  {Bishop}}, \bibinfo {author} {\bibfnamefont {P.~H.~Y.}\ \bibnamefont {Li.}},
  \bibinfo {author} {\bibfnamefont {D.~J.~J.}\ \bibnamefont {Farnell}}, \ and\
  \bibinfo {author} {\bibfnamefont {C.~E.}\ \bibnamefont {Campbell}},\
  }\href@noop {} {\bibfield  {journal} {\bibinfo  {journal} {J.\ Phys.:
  Condens.\ Matter}\ }\textbf {\bibinfo {volume} {24}},\ \bibinfo {pages}
  {236002} (\bibinfo {year} {2012})}\BibitemShut {NoStop}%
\bibitem [{\citenamefont {Lamas}, \citenamefont {Capponi},\ and\ \citenamefont
  {Pujol}(2011)}]{lamas-capponi}%
  \BibitemOpen
  \bibfield  {author} {\bibinfo {author} {\bibfnamefont {C.~A.}\ \bibnamefont
  {Lamas}}, \bibinfo {author} {\bibfnamefont {S.}~\bibnamefont {Capponi}}, \
  and\ \bibinfo {author} {\bibfnamefont {P.}~\bibnamefont {Pujol}},\ }\href
  {\doibase 10.1103/PhysRevB.84.115125} {\bibfield  {journal} {\bibinfo
  {journal} {Phys. Rev. B}\ }\textbf {\bibinfo {volume} {84}},\ \bibinfo
  {pages} {115125} (\bibinfo {year} {2011})}\BibitemShut {NoStop}%
\bibitem [{\citenamefont {Gong}\ \emph {et~al.}(2013)\citenamefont {Gong},
  \citenamefont {Sheng}, \citenamefont {Motrunich},\ and\ \citenamefont
  {Fisher}}]{Fisher_2013}%
  \BibitemOpen
  \bibfield  {author} {\bibinfo {author} {\bibfnamefont {S.-S.}\ \bibnamefont
  {Gong}}, \bibinfo {author} {\bibfnamefont {D.}~\bibnamefont {Sheng}},
  \bibinfo {author} {\bibfnamefont {O.~I.}\ \bibnamefont {Motrunich}}, \ and\
  \bibinfo {author} {\bibfnamefont {M.~P.}\ \bibnamefont {Fisher}},\
  }\href@noop {} {\bibfield  {journal} {\bibinfo  {journal} {Phys. Rev. B}\
  }\textbf {\bibinfo {volume} {88}},\ \bibinfo {pages} {165138} (\bibinfo
  {year} {2013})}\BibitemShut {NoStop}%
\bibitem [{\citenamefont {Cabra}, \citenamefont {Lamas},\ and\ \citenamefont
  {Rosales}(2011{\natexlab{a}})}]{Cabra_honeycomb_prb}%
  \BibitemOpen
  \bibfield  {author} {\bibinfo {author} {\bibfnamefont {D.~C.}\ \bibnamefont
  {Cabra}}, \bibinfo {author} {\bibfnamefont {C.~A.}\ \bibnamefont {Lamas}}, \
  and\ \bibinfo {author} {\bibfnamefont {H.~D.}\ \bibnamefont {Rosales}},\
  }\href@noop {} {\bibfield  {journal} {\bibinfo  {journal} {Phys. Rev. B}\
  }\textbf {\bibinfo {volume} {83}},\ \bibinfo {pages} {094506} (\bibinfo
  {year} {2011}{\natexlab{a}})}\BibitemShut {NoStop}%
\bibitem [{\citenamefont {Messio}, \citenamefont {Bernu},\ and\ \citenamefont
  {Lhuillier}(2012)}]{Messio}%
  \BibitemOpen
  \bibfield  {author} {\bibinfo {author} {\bibfnamefont {L.}~\bibnamefont
  {Messio}}, \bibinfo {author} {\bibfnamefont {B.}~\bibnamefont {Bernu}}, \
  and\ \bibinfo {author} {\bibfnamefont {C.}~\bibnamefont {Lhuillier}},\
  }\href@noop {} {\bibfield  {journal} {\bibinfo  {journal} {Phys. Rev. Lett.}\
  }\textbf {\bibinfo {volume} {108}},\ \bibinfo {pages} {207204} (\bibinfo
  {year} {2012})}\BibitemShut {NoStop}%
\bibitem [{\citenamefont {Honecker}, \citenamefont {Mila},\ and\ \citenamefont
  {Troyer}(2000)}]{Honecker2000a}%
  \BibitemOpen
  \bibfield  {author} {\bibinfo {author} {\bibfnamefont {A.}~\bibnamefont
  {Honecker}}, \bibinfo {author} {\bibfnamefont {F.}~\bibnamefont {Mila}}, \
  and\ \bibinfo {author} {\bibfnamefont {M.}~\bibnamefont {Troyer}},\ }\href
  {\doibase 10.1007/s100510051120} {\bibfield  {journal} {\bibinfo  {journal}
  {Eur. Phys. J. B}\ }\textbf {\bibinfo {volume} {15}},\ \bibinfo {pages} {227}
  (\bibinfo {year} {2000})}\BibitemShut {NoStop}%
\bibitem [{\citenamefont {Mosadeq}, \citenamefont {Shahbazi},\ and\
  \citenamefont {Jafari}(2011)}]{Mosadeq}%
  \BibitemOpen
  \bibfield  {author} {\bibinfo {author} {\bibfnamefont {H.}~\bibnamefont
  {Mosadeq}}, \bibinfo {author} {\bibfnamefont {F.}~\bibnamefont {Shahbazi}}, \
  and\ \bibinfo {author} {\bibfnamefont {S.}~\bibnamefont {Jafari}},\
  }\href@noop {} {\bibfield  {journal} {\bibinfo  {journal} {Journal of
  Physics: Condensed Matter}\ }\textbf {\bibinfo {volume} {23}},\ \bibinfo
  {pages} {226006} (\bibinfo {year} {2011})}\BibitemShut {NoStop}%
\bibitem [{\citenamefont {Mulder}\ \emph {et~al.}(2010)\citenamefont {Mulder},
  \citenamefont {Ganesh}, \citenamefont {Capriotti},\ and\ \citenamefont
  {Paramekanti}}]{Mulder}%
  \BibitemOpen
  \bibfield  {author} {\bibinfo {author} {\bibfnamefont {A.}~\bibnamefont
  {Mulder}}, \bibinfo {author} {\bibfnamefont {R.}~\bibnamefont {Ganesh}},
  \bibinfo {author} {\bibfnamefont {L.}~\bibnamefont {Capriotti}}, \ and\
  \bibinfo {author} {\bibfnamefont {A.}~\bibnamefont {Paramekanti}},\
  }\href@noop {} {\bibfield  {journal} {\bibinfo  {journal} {Phys. Rev. B}\
  }\textbf {\bibinfo {volume} {81}},\ \bibinfo {pages} {214419} (\bibinfo
  {year} {2010})}\BibitemShut {NoStop}%
\bibitem [{\citenamefont {Ceccatto}, \citenamefont {Gazza},\ and\ \citenamefont
  {Trumper}(1993)}]{Trumper1}%
  \BibitemOpen
  \bibfield  {author} {\bibinfo {author} {\bibfnamefont {H.}~\bibnamefont
  {Ceccatto}}, \bibinfo {author} {\bibfnamefont {C.}~\bibnamefont {Gazza}}, \
  and\ \bibinfo {author} {\bibfnamefont {A.}~\bibnamefont {Trumper}},\
  }\href@noop {} {\bibfield  {journal} {\bibinfo  {journal} {Phys. Rev. B}\
  }\textbf {\bibinfo {volume} {47}},\ \bibinfo {pages} {12329} (\bibinfo {year}
  {1993})}\BibitemShut {NoStop}%
\bibitem [{\citenamefont {Lamas}\ and\ \citenamefont
  {Matera}(2015)}]{Lamas2015b}%
  \BibitemOpen
  \bibfield  {author} {\bibinfo {author} {\bibfnamefont {C.~A.}\ \bibnamefont
  {Lamas}}\ and\ \bibinfo {author} {\bibfnamefont {J.~M.}\ \bibnamefont
  {Matera}},\ }\href@noop {} {\bibfield  {journal} {\bibinfo  {journal} {Phys.
  Rev. B}\ }\textbf {\bibinfo {volume} {92}},\ \bibinfo {pages} {115111}
  (\bibinfo {year} {2015})}\BibitemShut {NoStop}%
\bibitem [{\citenamefont {Brenig}\ and\ \citenamefont
  {Becker}(2001)}]{Brenig2001a}%
  \BibitemOpen
  \bibfield  {author} {\bibinfo {author} {\bibfnamefont {W.}~\bibnamefont
  {Brenig}}\ and\ \bibinfo {author} {\bibfnamefont {K.~W.}\ \bibnamefont
  {Becker}},\ }\href {\doibase 10.1103/PhysRevB.64.214413} {\bibfield
  {journal} {\bibinfo  {journal} {Phys. Rev. B}\ }\textbf {\bibinfo {volume}
  {64}},\ \bibinfo {pages} {214413} (\bibinfo {year} {2001})}\BibitemShut
  {NoStop}%
\bibitem [{\citenamefont {Chubukov}\ and\ \citenamefont
  {Jolicoeur}(1991)}]{Chubukov1991}%
  \BibitemOpen
  \bibfield  {author} {\bibinfo {author} {\bibfnamefont {A.~V.}\ \bibnamefont
  {Chubukov}}\ and\ \bibinfo {author} {\bibfnamefont {T.}~\bibnamefont
  {Jolicoeur}},\ }\href {\doibase 10.1103/PhysRevB.44.12050} {\bibfield
  {journal} {\bibinfo  {journal} {Phys. Rev. B}\ }\textbf {\bibinfo {volume}
  {44}},\ \bibinfo {pages} {12050(R)} (\bibinfo {year} {1991})}\BibitemShut
  {NoStop}%
\bibitem [{\citenamefont {Zhang}, \citenamefont {Arlego},\ and\ \citenamefont
  {Lamas}(2014)}]{Zhang2014}%
  \BibitemOpen
  \bibfield  {author} {\bibinfo {author} {\bibfnamefont {H.}~\bibnamefont
  {Zhang}}, \bibinfo {author} {\bibfnamefont {M.}~\bibnamefont {Arlego}}, \
  and\ \bibinfo {author} {\bibfnamefont {C.~A.}\ \bibnamefont {Lamas}},\ }\href
  {\doibase 10.1103/PhysRevB.89.024403} {\bibfield  {journal} {\bibinfo
  {journal} {Phys. Rev. B}\ }\textbf {\bibinfo {volume} {89}},\ \bibinfo
  {pages} {024403} (\bibinfo {year} {2014})}\BibitemShut {NoStop}%
\bibitem [{\citenamefont {Lamas}\ \emph {et~al.}(2013)\citenamefont {Lamas},
  \citenamefont {Ralko}, \citenamefont {Oshikawa}, \citenamefont {Poilblanc},\
  and\ \citenamefont {Pujol}}]{Oshikawa_2013}%
  \BibitemOpen
  \bibfield  {author} {\bibinfo {author} {\bibfnamefont {C.~A.}\ \bibnamefont
  {Lamas}}, \bibinfo {author} {\bibfnamefont {A.}~\bibnamefont {Ralko}},
  \bibinfo {author} {\bibfnamefont {M.}~\bibnamefont {Oshikawa}}, \bibinfo
  {author} {\bibfnamefont {D.}~\bibnamefont {Poilblanc}}, \ and\ \bibinfo
  {author} {\bibfnamefont {P.}~\bibnamefont {Pujol}},\ }\href@noop {}
  {\bibfield  {journal} {\bibinfo  {journal} {Physical Review B}\ }\textbf
  {\bibinfo {volume} {87}},\ \bibinfo {pages} {104512} (\bibinfo {year}
  {2013})}\BibitemShut {NoStop}%
\bibitem [{\citenamefont {Matsuda}\ \emph {et~al.}(2010)\citenamefont
  {Matsuda}, \citenamefont {Azuma}, \citenamefont {Tokunaga}, \citenamefont
  {Shimakawa},\ and\ \citenamefont {Kumada}}]{matsuda2010disordered}%
  \BibitemOpen
  \bibfield  {author} {\bibinfo {author} {\bibfnamefont {M.}~\bibnamefont
  {Matsuda}}, \bibinfo {author} {\bibfnamefont {M.}~\bibnamefont {Azuma}},
  \bibinfo {author} {\bibfnamefont {M.}~\bibnamefont {Tokunaga}}, \bibinfo
  {author} {\bibfnamefont {Y.}~\bibnamefont {Shimakawa}}, \ and\ \bibinfo
  {author} {\bibfnamefont {N.}~\bibnamefont {Kumada}},\ }\href@noop {}
  {\bibfield  {journal} {\bibinfo  {journal} {Physical review letters}\
  }\textbf {\bibinfo {volume} {105}},\ \bibinfo {pages} {187201} (\bibinfo
  {year} {2010})}\BibitemShut {NoStop}%
\bibitem [{\citenamefont {Smirnova}\ \emph {et~al.}(2009)\citenamefont
  {Smirnova}, \citenamefont {Azuma}, \citenamefont {Kumada}, \citenamefont
  {Kusano}, \citenamefont {Matsuda}, \citenamefont {Shimakawa}, \citenamefont
  {Takei}, \citenamefont {Yonesaki},\ and\ \citenamefont
  {Kinomura}}]{smirnova2009synthesis}%
  \BibitemOpen
  \bibfield  {author} {\bibinfo {author} {\bibfnamefont {O.}~\bibnamefont
  {Smirnova}}, \bibinfo {author} {\bibfnamefont {M.}~\bibnamefont {Azuma}},
  \bibinfo {author} {\bibfnamefont {N.}~\bibnamefont {Kumada}}, \bibinfo
  {author} {\bibfnamefont {Y.}~\bibnamefont {Kusano}}, \bibinfo {author}
  {\bibfnamefont {M.}~\bibnamefont {Matsuda}}, \bibinfo {author} {\bibfnamefont
  {Y.}~\bibnamefont {Shimakawa}}, \bibinfo {author} {\bibfnamefont
  {T.}~\bibnamefont {Takei}}, \bibinfo {author} {\bibfnamefont
  {Y.}~\bibnamefont {Yonesaki}}, \ and\ \bibinfo {author} {\bibfnamefont
  {N.}~\bibnamefont {Kinomura}},\ }\href@noop {} {\bibfield  {journal}
  {\bibinfo  {journal} {Journal of the American Chemical Society}\ }\textbf
  {\bibinfo {volume} {131}},\ \bibinfo {pages} {8313} (\bibinfo {year}
  {2009})}\BibitemShut {NoStop}%
\bibitem [{Note1()}]{Note1}%
  \BibitemOpen
  \bibinfo {note} {In DFT, the test wave functions are not exactly Slater's
  determinats, but the argument holds: GS associated to different spin
  configurations are orthogonal among them.}\BibitemShut {Stop}%
\bibitem [{\citenamefont {Sachdev}\ and\ \citenamefont
  {Bhatt}(1990)}]{Sachdev1990}%
  \BibitemOpen
  \bibfield  {author} {\bibinfo {author} {\bibfnamefont {S.}~\bibnamefont
  {Sachdev}}\ and\ \bibinfo {author} {\bibfnamefont {R.~N.}\ \bibnamefont
  {Bhatt}},\ }\href {\doibase 10.1103/PhysRevB.41.9323} {\bibfield  {journal}
  {\bibinfo  {journal} {Phys. Rev. B}\ }\textbf {\bibinfo {volume} {41}},\
  \bibinfo {pages} {9323} (\bibinfo {year} {1990})}\BibitemShut {NoStop}%
\bibitem [{\citenamefont {Trumper}\ \emph {et~al.}(1997)\citenamefont
  {Trumper}, \citenamefont {Manuel}, \citenamefont {Gazza},\ and\ \citenamefont
  {Ceccatto}}]{Trumper2}%
  \BibitemOpen
  \bibfield  {author} {\bibinfo {author} {\bibfnamefont {A.}~\bibnamefont
  {Trumper}}, \bibinfo {author} {\bibfnamefont {L.}~\bibnamefont {Manuel}},
  \bibinfo {author} {\bibfnamefont {C.}~\bibnamefont {Gazza}}, \ and\ \bibinfo
  {author} {\bibfnamefont {H.}~\bibnamefont {Ceccatto}},\ }\href@noop {}
  {\bibfield  {journal} {\bibinfo  {journal} {Phys. Rev. Lett.}\ }\textbf
  {\bibinfo {volume} {78}},\ \bibinfo {pages} {2216} (\bibinfo {year}
  {1997})}\BibitemShut {NoStop}%
\bibitem [{\citenamefont {Flint}\ and\ \citenamefont
  {Coleman}(2009)}]{Coleman}%
  \BibitemOpen
  \bibfield  {author} {\bibinfo {author} {\bibfnamefont {R.}~\bibnamefont
  {Flint}}\ and\ \bibinfo {author} {\bibfnamefont {P.}~\bibnamefont
  {Coleman}},\ }\href@noop {} {\bibfield  {journal} {\bibinfo  {journal}
  {Physical Review B}\ }\textbf {\bibinfo {volume} {79}},\ \bibinfo {pages}
  {014424} (\bibinfo {year} {2009})}\BibitemShut {NoStop}%
\bibitem [{\citenamefont {Zhang}\ and\ \citenamefont
  {Lamas}(2013)}]{Zhang_PRB_2013}%
  \BibitemOpen
  \bibfield  {author} {\bibinfo {author} {\bibfnamefont {H.}~\bibnamefont
  {Zhang}}\ and\ \bibinfo {author} {\bibfnamefont {C.}~\bibnamefont {Lamas}},\
  }\href@noop {} {\bibfield  {journal} {\bibinfo  {journal} {Phys. Rev. B}\
  }\textbf {\bibinfo {volume} {87}},\ \bibinfo {pages} {024415} (\bibinfo
  {year} {2013})}\BibitemShut {NoStop}%
\bibitem [{\citenamefont {Zhang}\ \emph {et~al.}(2016)\citenamefont {Zhang},
  \citenamefont {Lamas}, \citenamefont {Arlego},\ and\ \citenamefont
  {Brenig}}]{Brenig2016}%
  \BibitemOpen
  \bibfield  {author} {\bibinfo {author} {\bibfnamefont {H.}~\bibnamefont
  {Zhang}}, \bibinfo {author} {\bibfnamefont {C.~A.}\ \bibnamefont {Lamas}},
  \bibinfo {author} {\bibfnamefont {M.}~\bibnamefont {Arlego}}, \ and\ \bibinfo
  {author} {\bibfnamefont {W.}~\bibnamefont {Brenig}},\ }\href {\doibase
  10.1103/PhysRevB.93.235150} {\bibfield  {journal} {\bibinfo  {journal} {Phys.
  Rev. B}\ }\textbf {\bibinfo {volume} {93}},\ \bibinfo {pages} {235150}
  (\bibinfo {year} {2016})}\BibitemShut {NoStop}%
\bibitem [{\citenamefont {Lamas}\ \emph {et~al.}(2015)\citenamefont {Lamas},
  \citenamefont {Cabra}, \citenamefont {Pujol},\ and\ \citenamefont
  {Rossini}}]{Lamas2015}%
  \BibitemOpen
  \bibfield  {author} {\bibinfo {author} {\bibfnamefont {C.~A.}\ \bibnamefont
  {Lamas}}, \bibinfo {author} {\bibfnamefont {D.~C.}\ \bibnamefont {Cabra}},
  \bibinfo {author} {\bibfnamefont {P.}~\bibnamefont {Pujol}}, \ and\ \bibinfo
  {author} {\bibfnamefont {G.~L.}\ \bibnamefont {Rossini}},\ }\href {\doibase
  10.1140/epjb/e2015-60211-6} {\bibfield  {journal} {\bibinfo  {journal} {The
  European Physical Journal B}\ }\textbf {\bibinfo {volume} {88}},\ \bibinfo
  {pages} {176} (\bibinfo {year} {2015})}\BibitemShut {NoStop}%
\bibitem [{\citenamefont {Elias}, \citenamefont {Arlego},\ and\ \citenamefont
  {Lamas}(2017)}]{Elias2017}%
  \BibitemOpen
  \bibfield  {author} {\bibinfo {author} {\bibfnamefont {F.}~\bibnamefont
  {Elias}}, \bibinfo {author} {\bibfnamefont {M.}~\bibnamefont {Arlego}}, \
  and\ \bibinfo {author} {\bibfnamefont {C.~A.}\ \bibnamefont {Lamas}},\ }\href
  {\doibase 10.1103/PhysRevB.95.214426} {\bibfield  {journal} {\bibinfo
  {journal} {Phys. Rev. B}\ }\textbf {\bibinfo {volume} {95}},\ \bibinfo
  {pages} {214426} (\bibinfo {year} {2017})}\BibitemShut {NoStop}%
\bibitem [{\citenamefont {Bhatia}(1997)}]{Bhatia97}%
  \BibitemOpen
  \bibfield  {author} {\bibinfo {author} {\bibfnamefont {R.}~\bibnamefont
  {Bhatia}},\ }\href@noop {} {\emph {\bibinfo {title} {Matrix Analysis}}},\
  Vol.\ \bibinfo {volume} {169}\ (\bibinfo  {publisher} {Springer},\ \bibinfo
  {year} {1997})\BibitemShut {NoStop}%
\bibitem [{\citenamefont {Matera}\ and\ \citenamefont
  {Lamas}(2018)}]{visualbond}%
  \BibitemOpen
  \bibfield  {author} {\bibinfo {author} {\bibfnamefont {J.~M.}\ \bibnamefont
  {Matera}}\ and\ \bibinfo {author} {\bibfnamefont {C.~A.}\ \bibnamefont
  {Lamas}},\ }\href@noop {} {\enquote {\bibinfo {title} {Visualbond
  spectrojotometer},}\ }\bibinfo {howpublished} {available in
  \url{https://github.com/mmatera/spectrojotometer}} (\bibinfo {year}
  {2018})\BibitemShut {NoStop}%
\bibitem [{\citenamefont {Arlego}, \citenamefont {Lamas},\ and\ \citenamefont
  {Zhang}(2014)}]{Arlego201415}%
  \BibitemOpen
  \bibfield  {author} {\bibinfo {author} {\bibfnamefont {M.}~\bibnamefont
  {Arlego}}, \bibinfo {author} {\bibfnamefont {C.~A.}\ \bibnamefont {Lamas}}, \
  and\ \bibinfo {author} {\bibfnamefont {H.}~\bibnamefont {Zhang}},\ }\href
  {\doibase 10.1088/1742-6596/568/4/042019} {\bibfield  {journal} {\bibinfo
  {journal} {J. Phys.: Conf. Ser.}\ }\textbf {\bibinfo {volume} {568}},\
  \bibinfo {pages} {042019} (\bibinfo {year} {2014})}\BibitemShut {NoStop}%
\bibitem [{\citenamefont {Cabra}, \citenamefont {Lamas},\ and\ \citenamefont
  {Rosales}(2011{\natexlab{b}})}]{Cabra_honeycomb_2}%
  \BibitemOpen
  \bibfield  {author} {\bibinfo {author} {\bibfnamefont {D.}~\bibnamefont
  {Cabra}}, \bibinfo {author} {\bibfnamefont {C.}~\bibnamefont {Lamas}}, \ and\
  \bibinfo {author} {\bibfnamefont {H.}~\bibnamefont {Rosales}},\ }\href@noop
  {} {\bibfield  {journal} {\bibinfo  {journal} {Mod. Phys. Lett. B}\ }\textbf
  {\bibinfo {volume} {25}},\ \bibinfo {pages} {891} (\bibinfo {year}
  {2011}{\natexlab{b}})}\BibitemShut {NoStop}%
\bibitem [{\citenamefont {Fouet}, \citenamefont {Sindzingre},\ and\
  \citenamefont {Lhuillier}(1975)}]{fouet1975investigation}%
  \BibitemOpen
  \bibfield  {author} {\bibinfo {author} {\bibfnamefont {J.}~\bibnamefont
  {Fouet}}, \bibinfo {author} {\bibfnamefont {P.}~\bibnamefont {Sindzingre}}, \
  and\ \bibinfo {author} {\bibfnamefont {C.}~\bibnamefont {Lhuillier}},\
  }\href@noop {} {\bibfield  {journal} {\bibinfo  {journal} {Zeitschrift
  f{\"u}r Physik B Condensed Matter}\ }\textbf {\bibinfo {volume} {20}},\
  \bibinfo {pages} {241} (\bibinfo {year} {1975})}\BibitemShut {NoStop}%
\bibitem [{\citenamefont {Kandpal}\ and\ \citenamefont {van~den
  Brink}(2011)}]{kandpal2011calculation}%
  \BibitemOpen
  \bibfield  {author} {\bibinfo {author} {\bibfnamefont {H.~C.}\ \bibnamefont
  {Kandpal}}\ and\ \bibinfo {author} {\bibfnamefont {J.}~\bibnamefont {van~den
  Brink}},\ }\href@noop {} {\bibfield  {journal} {\bibinfo  {journal} {Physical
  Review B}\ }\textbf {\bibinfo {volume} {83}},\ \bibinfo {pages} {140412}
  (\bibinfo {year} {2011})}\BibitemShut {NoStop}%
\bibitem [{\citenamefont {Krokhmalskii}\ \emph {et~al.}(2017)\citenamefont
  {Krokhmalskii}, \citenamefont {Baliha}, \citenamefont {Derzhko},
  \citenamefont {Schulenburg},\ and\ \citenamefont {Richter}}]{Richter2017}%
  \BibitemOpen
  \bibfield  {author} {\bibinfo {author} {\bibfnamefont {T.}~\bibnamefont
  {Krokhmalskii}}, \bibinfo {author} {\bibfnamefont {V.}~\bibnamefont
  {Baliha}}, \bibinfo {author} {\bibfnamefont {O.}~\bibnamefont {Derzhko}},
  \bibinfo {author} {\bibfnamefont {J.}~\bibnamefont {Schulenburg}}, \ and\
  \bibinfo {author} {\bibfnamefont {J.}~\bibnamefont {Richter}},\ }\href
  {\doibase 10.1103/PhysRevB.95.094419} {\bibfield  {journal} {\bibinfo
  {journal} {Phys. Rev. B}\ }\textbf {\bibinfo {volume} {95}},\ \bibinfo
  {pages} {094419} (\bibinfo {year} {2017})}\BibitemShut {NoStop}%
\bibitem [{\citenamefont {Oitmaa}\ and\ \citenamefont
  {Singh}(2012)}]{Oitmaa_2012}%
  \BibitemOpen
  \bibfield  {author} {\bibinfo {author} {\bibfnamefont {J.}~\bibnamefont
  {Oitmaa}}\ and\ \bibinfo {author} {\bibfnamefont {R.}~\bibnamefont {Singh}},\
  }\href@noop {} {\bibfield  {journal} {\bibinfo  {journal} {Phys. Rev. B}\
  }\textbf {\bibinfo {volume} {85}},\ \bibinfo {pages} {014428} (\bibinfo
  {year} {2012})}\BibitemShut {NoStop}%
\bibitem [{\citenamefont {Okubo}\ \emph {et~al.}(2010)\citenamefont {Okubo},
  \citenamefont {Elmasry}, \citenamefont {Zhang}, \citenamefont {Fujisawa},
  \citenamefont {Sakurai}, \citenamefont {Ohta}, \citenamefont {Azuma},
  \citenamefont {Sumirnova},\ and\ \citenamefont {Kumada}}]{okubo2010high}%
  \BibitemOpen
  \bibfield  {author} {\bibinfo {author} {\bibfnamefont {S.}~\bibnamefont
  {Okubo}}, \bibinfo {author} {\bibfnamefont {F.}~\bibnamefont {Elmasry}},
  \bibinfo {author} {\bibfnamefont {W.}~\bibnamefont {Zhang}}, \bibinfo
  {author} {\bibfnamefont {M.}~\bibnamefont {Fujisawa}}, \bibinfo {author}
  {\bibfnamefont {T.}~\bibnamefont {Sakurai}}, \bibinfo {author} {\bibfnamefont
  {H.}~\bibnamefont {Ohta}}, \bibinfo {author} {\bibfnamefont {M.}~\bibnamefont
  {Azuma}}, \bibinfo {author} {\bibfnamefont {O.~A.}\ \bibnamefont
  {Sumirnova}}, \ and\ \bibinfo {author} {\bibfnamefont {N.}~\bibnamefont
  {Kumada}},\ }in\ \href@noop {} {\emph {\bibinfo {booktitle} {Journal of
  Physics: Conference Series}}},\ Vol.\ \bibinfo {volume} {200}\ (\bibinfo
  {organization} {IOP Publishing},\ \bibinfo {year} {2010})\ p.\ \bibinfo
  {pages} {022042}\BibitemShut {NoStop}%
\bibitem [{\citenamefont {Azuma}\ \emph {et~al.}(2011)\citenamefont {Azuma},
  \citenamefont {Matsuda}, \citenamefont {Onishi}, \citenamefont {Olga},
  \citenamefont {Kusano}, \citenamefont {Tokunaga}, \citenamefont {Shimakawa},\
  and\ \citenamefont {Kumada}}]{azuma2011frustrated}%
  \BibitemOpen
  \bibfield  {author} {\bibinfo {author} {\bibfnamefont {M.}~\bibnamefont
  {Azuma}}, \bibinfo {author} {\bibfnamefont {M.}~\bibnamefont {Matsuda}},
  \bibinfo {author} {\bibfnamefont {N.}~\bibnamefont {Onishi}}, \bibinfo
  {author} {\bibfnamefont {S.}~\bibnamefont {Olga}}, \bibinfo {author}
  {\bibfnamefont {Y.}~\bibnamefont {Kusano}}, \bibinfo {author} {\bibfnamefont
  {M.}~\bibnamefont {Tokunaga}}, \bibinfo {author} {\bibfnamefont
  {Y.}~\bibnamefont {Shimakawa}}, \ and\ \bibinfo {author} {\bibfnamefont
  {N.}~\bibnamefont {Kumada}},\ }in\ \href@noop {} {\emph {\bibinfo {booktitle}
  {Journal of Physics: Conference Series}}},\ Vol.\ \bibinfo {volume} {320}\
  (\bibinfo {organization} {IOP Publishing},\ \bibinfo {year} {2011})\ p.\
  \bibinfo {pages} {012005}\BibitemShut {NoStop}%
\bibitem [{\citenamefont {Alaei}\ \emph {et~al.}(2017)\citenamefont {Alaei},
  \citenamefont {Mosadeq}, \citenamefont {Sarsari},\ and\ \citenamefont
  {Shahbazi}}]{Alaei.2017}%
  \BibitemOpen
  \bibfield  {author} {\bibinfo {author} {\bibfnamefont {M.}~\bibnamefont
  {Alaei}}, \bibinfo {author} {\bibfnamefont {H.}~\bibnamefont {Mosadeq}},
  \bibinfo {author} {\bibfnamefont {I.~A.}\ \bibnamefont {Sarsari}}, \ and\
  \bibinfo {author} {\bibfnamefont {F.}~\bibnamefont {Shahbazi}},\ }\href
  {\doibase 10.1103/PhysRevB.96.140404} {\bibfield  {journal} {\bibinfo
  {journal} {Phys. Rev. B}\ }\textbf {\bibinfo {volume} {96}},\ \bibinfo
  {pages} {140404} (\bibinfo {year} {2017})}\BibitemShut {NoStop}%
\end{thebibliography}%

\end{document}